\documentclass[twocolumn,oneside,slac_two]{revtex4}
\usepackage{graphicx}
\usepackage{fancyhdr}
\usepackage{amsmath}
\usepackage{amsfonts}
\usepackage{amssymb}
\usepackage{amsthm}
\usepackage{amscd}

\usepackage[colorlinks=true]{hyperref}

\usepackage{latexsym}
\usepackage{mathrsfs}
\usepackage{psfrag}
\usepackage[dvips]{epsfig}
\usepackage{epsfig}
\usepackage[all]{xy}         

\newcommand{\R}{\mathbb{R}}                     
\newcommand{\K}{\mathbb{K}}

 \newcommand{\unit}{\mathbb{I}}
 \newcommand{\amat}{\mathbb{A}}
 \newcommand{\source}{$\mathcal{S}$}
 \newcommand{\detector}{$\mathcal{D}$}
 \newcommand{\causalset}{$\mathcal{C}$}
 \newcommand{\Q}{\mathbb{Q}}

 \newcommand{\CS}{CS}
  \newcommand{\CSs}{CSs}

\providecommand{\norm}[1]{\left\lVert#1\right\rVert}       

\providecommand{\abs}[1]{\left\lvert#1\right\rvert}        

\begin{document}

\title{Photon dispersion in causal sets}

%

\author{Jeffrey D. Scargle}
\affiliation{NASA Ames Research Center, Moffett Field, CA 94035-1000}
\author{Slobodan N. Simi\'c}
\affiliation{Department of Mathematics, San Jos\'e State University, San
  Jos\'e, CA 95192-0103}

\begin{abstract}
A very small dispersion
in the speed of light
may be observable in Fermi
time- and energy-tagged data
on variable sources, such as
gamma-ray bursts (GRB) and active galactic 
nuclei (AGN).
We describe a method to compute 
the size of this effect by
applying the 
Feynman sum-over-histories
formalism for relativistic quantum electrodynamics
to a discrete model of 
space-time called \emph{causal set theory}.
\end{abstract}

\maketitle

\thispagestyle{fancy}


\section{Introduction}               \label{sec:intro}

The vacuum speed of light 
may be a function of photon energy.
Even a very small such dependence 
might yield energy-dependent time delays which,
over the long journeys from distant astronomical sources,
may accumulate to a level measurable
by the Fermi Gamma Ray Space Telescope~\cite{abdo}. 
The expectation from general considerations
\cite{gac}
is that the scale of the effect is 
linear in the photon's energy:
\begin{equation}
{dt \over T} \sim { dE \over E_{Planck} },
\end{equation}
\noindent
where the Planck energy is
\begin{equation}
E_{Planck} = \sqrt{ \hbar c ^{5} \over G  }
\sim 1.22 \times 10^{19} GeV \ .
\end{equation}
\noindent
This relation gives delays over astronomical
distances of seconds (GeV energies) 
to hours (TeV energies).

A motivation for this
suggestion is the possibility
that space-time may be lumpy,
fuzzy, or even discrete, on small
scales, and that this lumpiness 
may affect photon propagation in
a dispersive way.
We adopt the extreme viewpoint
provided by 
\emph{causal set theory},
an approach to quantum gravity 
pioneered by Rafael Sorkin
\cite{sorkin+90},
with the founding principle 
that spacetime is a \emph{partially ordered} 
set of purely discrete points.
The partial order nicely reflects the 
causal relationships between events
fundamental to special relativity.
We also adopt the Feynman
sum-over-histories quantum
electrodynamics, 
which is greatly simplified 
by the postulated discreteness:
path integrals, and the difficulties
associated with divergences
and defining the measure on
the space of all paths
are replaced by discrete sums.

\section{Causal Sets}

Causal set (CS) theory postulates that at very small scales -- below
those of ordinary experience, and even below the smallest scales
accessible to particle physics experiments -- space-time consists of a
set of discrete points.  The usual space-time continuum is thus
regarded as a macroscopic construct that does not exist.  One only
requires that the large scale limit of the \CS \ recovers known,
macroscopic, continuum physics.

In mathematical terms, 
a \emph{causal set} is a pair ($\mathcal{C},\prec$), where 
\causalset \  is a set and $\prec$ is a binary relation 
on \causalset  \ satisfying the following properties
($x$, $y$, $z$ etc. are general points in \causalset):
\begin{enumerate}
\item {\bf Partial Order: } Some, but not all, pairs $x,y$ of points are ordered: $x \prec y$;
\item {\bf Transitivity:} If $x \prec y$ and $y \prec z$, then $x \prec z$;
\item {\bf Anti-reflexivity:} No point is related to itself: $x \not\prec x$;
\item {\bf Local finiteness:} For all $x, y \in $ \causalset, 
the set $[x,y] = \{ z \in $ \causalset $ : x \prec z \ \ \text{and} \ \ z \prec y \}$ is finite.
\end{enumerate}
\noindent
The first two items are consistent with,
and are taken to express,
the causal relations
of special relativity: $x \prec y$ (read ``$x$ precedes $y$'') means that
$y$ is in the forward light-cone of $x$, and can affect $x$.  
If separated by a space-like interval
$x$ and $y$ are not ordered.
Transitivity here simply expresses
the relation between nested light cones.

One consequence of the above ansatz 
is that all space-time information 
is contained in the causal connections
among the points.
For example, any topological
or metric notions must be extracted from
the discrete (i.e., combinatorial) structure of (\causalset, $\prec)$
\cite{henson_2,ilie,major,rideout}, 
without any reference to a 
continuum.
This also means that the action (see below)
for the step $x \rightarrow y$
can depend on only the causal 
relation between the two points,
plus the energy and polarization
of the photon. 
We make use of this simplifying fact below.

Note that arbitrarily defined partial orderings among the points in
general can yield causal sets that cannot be realized as a subset of
any Lorentzian manifold. The \emph{sprinkling} procedure
\cite{henson+06} for generating \CS \ order relations was invented to
avoid these microscopic inconsistencies, to avoid other problems that
might foul up the macroscopic continuum limit\footnote{The viewpoint
  here is that the \CS \ is the reality, and the continuum is the
  approximation to it; this is the opposite of the usual view.}
mentioned above, and to ensure a kind of statistical Lorentz
invariance.  The process begins by randomly sprinkling points in a
continuous Lorentzian space-time manifold $M$, of the appropriate
dimension.  For example, to model flat space, the points would be
uniformly and independently distributed\footnote{This means that the
  probability of sprinkling $n$ points is a region of volume $V$ is
  $(\varrho V)^n e^{-\varrho V}/n!$.  Hence this is often called a
  Poisson process, but the essence is the independent and uniform
  nature of the random distribution, not the Poisson distribution
  function of counts in cells.} in $M$.  The fundamental density
$\varrho$ of this distribution is frequently taken to be such that the
mean (continuum) separation of points is the Planck scale.  Then the
partial order relations among the sprinkled points are derived from
the causal (light-cone) relations assessed in the continuum $M$.  Here
we adopt this sprinkling concept as defining causal sets, which in the
literature are said to be \emph{embedded} into the continuum $M$.

\section{Photon Propagation: Feynman Sums Over Space-Time Paths }   \label{sec:feynman}

Given a causal set (\causalset, $\prec)$ embedded 
in some $M = \R^{d+1}$, 
where $d$ is the spatial dimension
and there is one time dimension,
we consider photon 
propagation in \causalset. 
Given spacetime points $x, y$, one asks for the value of the
\emph{propagator} $\K(x,y)$, \emph{i.e.} the quantum-mechanical
amplitude for a photon emitted at $x$ to be detected at $y$; the
probability of this transtion is $\abs{\K(x,y)}^2$.  The standard sum
over paths formalism\cite{feynman_1,feynman_2} states that the
amplitude for each path from $x$ to $y$ is the product of the
amplitudes of each step comprising the path, and the amplitude for the
process is obtained by summing these amplitudes over {\bf all paths}
from $x$ to $y$.

It is more realistic to take \source,
a subset of \causalset, to represent
an unresolved astronomical source, 
and another subset \detector  \ to 
represent the detector.
Then the usual prescription is that the 
amplitude for the process 
of interest -- photons emitted from the source
arriving at the detector, symbolically 
\source $ \rightarrow $ \detector \ \ -- is the sum of the
amplitudes over all paths starting
beginning at some point in \source  \ and
ending at some point in \detector.
The probability of the process \source $ \rightarrow $ \detector \ 
is the square of the complex amplitude
of the process.
This prescription is valid if and only
if one is not able to determine which
point in \detector \  the photon arrives at, 
nor from which point in \source \  the photon
originates.  
If the detector were to be
divided into \emph{pixels}, 
such that the identity of the
pixel at which the photon arrives
is knowable,
then as usual 
amplitudes are additive
within pixels, but
probabilities are
additive across pixels.


To compute the propagator $\K(x,y)$ we use the Feynman path integral
approach which asserts that the propagator is the ``integral'' of
$\exp( i S(\gamma)/\hbar)$ over the space of all continuous paths
$\gamma$ from $x$ to $y$, where $S(\gamma)$ is the action associated
with $\gamma$. Whereas in the continuous case this integral is not
well-defined in a rigorous sense and its meaning, definition and
computation are still an area of investigation, in the causal set
scenario the integral reduces to a sum, to which it is easier to give
a precise meaning. Related work can be found in
\cite{johnston_1,johnston_2}, but unlike these references our approach
takes the \emph{all paths} spirit of \cite{feynman_2} quite literally,
allowing \textit{causal, non-causal, and superluminal paths} and
relying on the phase averages for large action paths as appropriate
weights.

All that remains is specification of the
action $S(x,y)$ for an elementary step 
$x \rightarrow y$, for arbitrary points
$x$ and $y$ in \causalset,
for the corresponding amplitude is~\cite{feynman_1}
\begin{equation}   \label{eq:action}
S(x,y) =  S_{0} e^{ i { S(x,y) \over \hbar }  } 
\end{equation}
where $\hbar$ is Planck's constant, and 
$S_{0}$ is a normalization factor
which can be determined by considering 
the operator for an infinitesimal 
advance in time (\cite{feynman_1}, \S 6).

\section{Mathematical results}        \label{sec:math-results}

To effect a suitable choice of the action 
just described, take the amplitude 
of each causal or non-causal jump
to be $a$ or $b$, respectively,
where $a$ and $b$ are 
suitably chosen complex numbers. 
Given a path $\gamma = x_0 x_1 \cdots x_n$ 
consisting of $n$ steps,
set $S(\gamma) = a^{n-k} b^k$, where $k$ is the number 
of non-causal steps in $\gamma$,
and $n-k$ the number of causal ones.
The propagator from $x \equiv x_0$ to $y \equiv x_n$ is then
\begin{equation}       \label{eq:K}
  \K(x,y) = \sum_{n=1}^\infty \sum_{k=0}^n a^{n-k} b^k p_{n,k}(x,y),
\end{equation}
where $p_{n,k}(x,y)$ denotes the number of 
paths of length $n$ from $x$ to $y$ 
with exactly $k$ non-causal steps.
We now adopt a statistical view and,
in a slight abuse of notation, treat
 $\K$ and $p_{n,k}(x,y)$ 
as averages over random sprinklings.

One obvious problem with this definition is that for $n \geq 2 $ and
$k \geq 1$, $p_{n,k}(x,y)$ is infinite. To rectify this problem, we
pick a sufficiently large number $r > 0$ and allow jumps (both casual
and non-causal) only up to distance $r$.  That is, $\gamma = x_0 x_1
\cdots x_n$ is defined to be an \emph{admissible path} only if each
step satisfies $d(x_j,x_{j+1}) \leq r$, where $d$ is the distance
metric.  This distance should be defined in the \CS \ sense
(\emph{e.g.}  as in \cite{rideout}), but in this preliminary study we
have ignored this nuance.  If we denote the corresponding propagator
by $\K_r$, then
\begin{equation}                \label{eq:Kr}
    \K_r(x,y) = \sum_{n=1}^\infty \sum_{k=0}^n a^{n-k} b^k p_{n,k}^r(x,y),
\end{equation}
where $p_{n,k}^r(x,y)$ is the expected number of admissible paths of
length $n$ from $x$ to $y$ with exactly $k$ non-causal steps [we will
call such paths $(n,k)$-paths].  One can think of the sum in equation
\eqref{eq:Kr} either as an ordinary series (whose convergence will be
assured if $\abs{a}$ and $\abs{b}$ are sufficiently small) or as an
\textit{oscillatory} series, whose convergence is defined
differently. We choose the former interpretation of equation
\eqref{eq:Kr}.

\begin{figure}[h]
\centerline{\includegraphics[width=0.7\hsize]{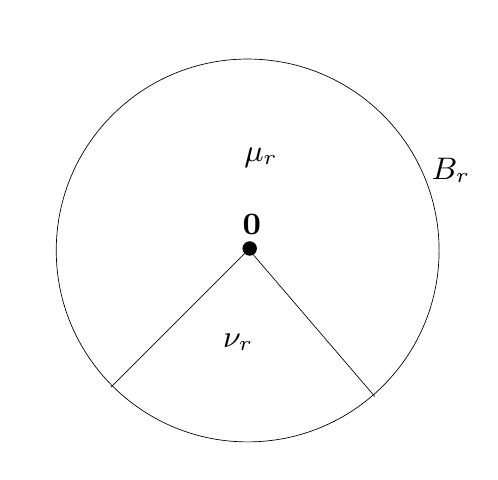}}
\caption{The region marked $\nu_r$ is the
backward light cone of $\mathbf{0}$, truncated
at distance $r$. The remaining region
marked $\mu_r$ includes the forward
light cone and the remaining region
with space-like separations from,
and therefore causally disconnected
from, $\mathbf{0}$.}
\label{fig:ball}
\end{figure}

\begin{figure}[htbp]
\centerline{\includegraphics[width=0.8\hsize]{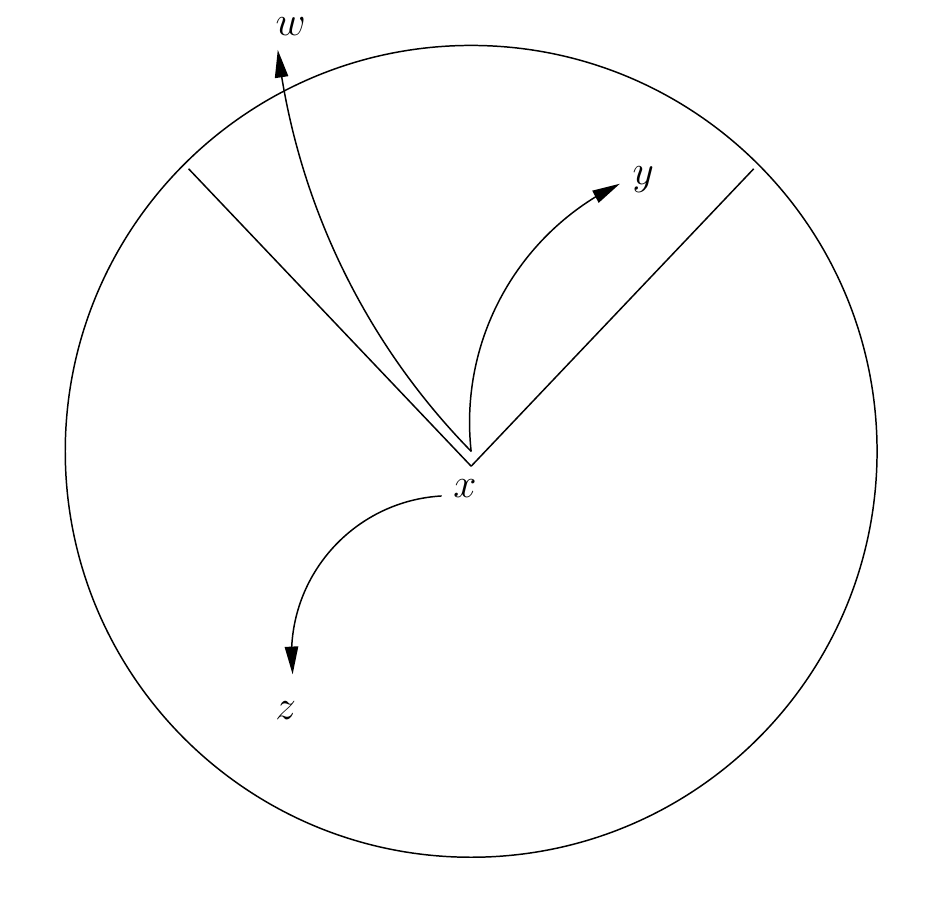}}
\caption{Jumps from $x$ to $y$ and $x$ to $z$ are admissible, but not
  from $x$ to $w$.}
\label{fig:jump}
\end{figure}

To compute $\K_r(x,y)$ for any $x, y \in \R^{d+1}$, we need a
suitable expression for $p_{n,k}^r(x,y)$. 
The amplitudes of causal and non-causal steps
will be represented by defining two functions $\nu_r,
\mu_r : \R^{d+1} \to \R$ as follows:
\begin{equation}          \label{eq:nu}
  \nu_r(z) = 
  \begin{cases}
    1 & \text{if} \ z \prec \mathbf{0} \ \text{and} \ d(z,\mathbf{0}) \leq r \\
    0 & \text{otherwise},
  \end{cases}
\end{equation}
and
\begin{equation}          \label{eq:mu}
  \mu_r(z) = 
  \begin{cases}
  1 & \text{if} \ z \not\prec \mathbf{0} \ \text{and} \ 
     d(z,\mathbf{0}) \leq r \\
  0 & \text{otherwise}.
  \end{cases}
\end{equation}
See Figure~\ref{fig:ball}. Observe that $\nu_r(x-y) = 1$ if and only
if there is an admissible non-causal jump (i.e., path of length one)
from $x$ to $y$, and $\mu_r(x-y) = 1$ if and only if there is an
admissible non-causal jump from $x$ to $y$.  It is not hard to see
that $\nu_r = \mathbf{1}_{B_r} \mathbf{1}_{L_-}$ and $\mu_r
=\mathbf{1}_{B_r} \mathbf{1}_{L_+}$ where $\mathbf{1}_{B_r}$ is the
characteristic function of the ball $B_r$ of radius $r$ centered at
the origin, and $\mathbf{1}_{L_{\pm}}$ are the characteristic
functions of the future and past light cones of the origin. See
Figure~\ref{fig:jump}.

It is possible to derive explicit formulas for $p_{n,k}^r(x,y)$ using
convolution-type integrals of functions $\nu_r, \mu_r$. For lack of
space, we do not present these formulas. It can then be shown that the
functions $p_{n,k}^r$ satisfy the following family of integral
equations:
\begin{widetext}

\begin{equation}         \label{eq:p_int}
  p_{n+1,k}^r(x,y) = \varrho \int_{\R^{d+1}} p_{n,k}^r(x,z) \nu_r(z-y) \: dz 
       + \varrho \int_{\R^{d+1}} p_{n,k-1}^r(x,z) \mu_r(z-y) \: dz,
\end{equation}
for all $n \geq 1$ and $1 \leq k \leq n$.
A lengthy calculation yields the following integral equation for the propagator:
\begin{align}         \label{eq:K_int}
  a \varrho \int_{\R^{d+1}} K_r(x,z) \nu_r(z-y) \: dz  & + b \varrho \int_{\R^{d+1}}
  \K_r(x,z) \mu_r(z-y) \: dz  = \K_r(x,y) - a \nu_r(x-y) - b \mu_r(x-y) \\
  & - \sum_{n=2}^\infty b^n p_{n,n}^r(x,y) + b \varrho \sum_{n=1}^\infty b^n
  \int_{\R^{d+1}} p_{n,n}^r(x,z) \mu_r(z-y) \: dz. \nonumber
\end{align}
\end{widetext}

To solve equation \eqref{eq:K_int}, we observe that both $\K_r(x,y)$
and $p_{n,n}^r(x,y)$ are translation invariant. This means that
$\K_r(x,y) = \Psi_r(x-y)$ and $p_{n,n}^r(x,y) = P_n^r(x-y)$, for some
functions $\Psi_r, P_n^r \in L^1(\R^{d+1})$. Note also that the
integrals in equation \eqref{eq:K_int} are convolutions. Applying the
Fourier transform $\widehat{\cdot}$, solving for $\widehat{\Psi}_r$,
and applying the inverse Fourier transform yields:
\begin{equation}     \label{eq:Psi}
  \Psi_r = \sum_{n=0}^\infty \varrho^n (a \nu_r + b \mu_r)^{\ast (n+1)},
\end{equation}
where $(a \nu_r + b \mu_r)^{\ast (n+1)}$ denotes the $(n+1)$-st
convolution power of $a \nu_r + b \mu_r$. To compute the geometric
convolution sum in equation \eqref{eq:Psi}, we need to recall that the
space $L^1(\R^{d+1})$ of Lebesgue integrable functions is a Banach
algebra \cite{rudin+fa} with respect to multiplication given by
convolution $\ast$, but that it does not possess a unit
element. However, a unity -- usually denoted by $\delta$ (one can
think of it as the Dirac delta ``function'') -- can be adjoined to
$L^1(\R^{d+1})$ to obtain a new Banach algebra $\tilde{L}^1(\R^{d+1})$
with unity (cf., \cite{rudin+fa}, \S 10.1). Then the sum in equation
\eqref{eq:Psi} becomes
\begin{equation}      \label{eq:Psi_sum}
  \Psi_r = (a \nu_r + b \mu_r) \ast 
  \{ \delta - \varrho(a \nu_r + b \mu_r) \}^{-1},
\end{equation}
where the inverse is taken in the extended algebra $\tilde{L}^1(\R^{d+1})$.

Now let \source \ and \detector \ be subsets of $\R^{d+1}$ representing
the source and the detector (see Section~\ref{sec:feynman}). Then the
probability a photon traveling from \source \ to \detector \ is
\begin{equation}    \label{eq:prob}
  \text{Prob}(\mathcal{S} \to \mathcal{D}) 
  = \abs{\int_{\mathcal{S} \times \mathcal{D}} \K_r(x,y) \: dx dy}^2.
\end{equation}
We leave the discussion of this integral for various \source \ and
\detector \ to our forthcoming longer paper.

\section{Numerical Results}

A matrix formalism simplifies 
numerical evaluation of the propagators
for small \CSs.  For a causal set
consisting of $N$ points $x_{n}, n = 1, 2, \dots N$,
define the amplitude matrix
\begin{equation}
\amat = [\amat_{n,m}] = [A(x_{n}, x_{m} )] \  ,
\end{equation}
generalizing the connection matrix
of graph theory by replacing the 
$1-0$ indicator of connectivity 
with the full amplitude from eq. \eqref{eq:action} for the
step between the points.
The product of  $\amat$
with itself,
\begin{equation}
[\amat^2]_{n,m} = \sum_{j} A_{n,j} A_{j,m} \ ,
\end{equation}
gives the amplitude for two-step
transitions $n \rightarrow m$
with a single intermediate stop at $j$.
Similarly $[\amat^k]_{n,m}$ gives the
amplitude of the same transition
with $k-1$ intermediate stops,
and the series
\begin{equation}
\Q = \amat + \amat^{2} + \amat^{3} + \dots 
\end{equation}
gives the amplitude summed over all
paths.
As long as $\norm \amat < 1$
for some norm, 
this series converges to the computationally useful form
\begin{equation}
\Q = ( \unit - \amat )^{-1} - \unit \ ,
\end{equation}
where $\unit$ is the indentity matrix.

So the amplitude of the transition 
\source \ $ \rightarrow $ \detector\ is
the sum of the elements of $\Q$
in the sub-block 
with column indices labeling points in \source\ 
and row indices labeling points in \detector,
and
the final probability for 
\source\  $ \rightarrow $ \detector\ is
the absolute square of this sum.

Considering this result and
the comments about pixels
which identify detected photons
made above, we see 
how precisely and simply 
these matrix operations implement 
the fundamental 
sum-over-histories quantum rules:
multiply amplitudes of steps on a path;
add amplitudes of all paths
between indistinguishable 
starting and final states;
add probabilities corresponding to
distinguishable final states.

After this construct was developed during a CAMCOS project at San
Jos\'e State University we discovered that Steven Johnston has a
similar approach~\cite{johnston_1}.

The formalism developed here seems to be a promising approach to
computing photon dispersion in the context of causal set theory.  A
future publication will elaborate the analytical approach in Section
IV and the numerical approach in Section V, including results on the
speed of light as a function of energy.

\bigskip 
\begin{acknowledgments}
  The authors wish to thank David Mattingly, Steven Johnston, Robert
  Israel, Harald Hanche-Olsen, and Rafael Sorkin for valuable
  comments, and Ricky Fernandez and students in the CAMCOS program at
  SJSU for various contributions.
\end{acknowledgments}

\bigskip 

\end{document}